\documentclass[twocolumn,showpacs,preprintnumbers,amsmath,amssymb]{revtex4}
\usepackage{dcolumn}
\usepackage{graphicx,amssymb,amsmath}

\begin{document}

\title{Central limit theorem for anomalous scaling due to correlations} 

\author{Fulvio Baldovin and Attilio L. Stella}
\email{baldovin@pd.infn.it, stella@pd.infn.it}
\affiliation{
Dipartimento di Fisica and
Sezione INFN, Universit\`a di Padova,\\
\it Via Marzolo 8, I-35131 Padova, Italy\\
}

\date{\today}

\begin{abstract}
We derive a central limit theorem for the probability distribution of
the sum of many critically correlated random variables. 
The theorem characterizes a variety of different processes sharing the
same asymptotic form of anomalous scaling and is based on a
correspondence with the  L\'evy-Gnedenko uncorrelated case.
In particular, correlated anomalous diffusion is mapped onto L\'evy diffusion. 
Under suitable assumptions, the nonstandard multiplicative structure
used for constructing the characteristic function of the total sum
allows us to determine correlations of  
partial sums exclusively on the basis of the global anomalous scaling.
\end{abstract}

\pacs{02.50.-r, 05.40.Fb, 89.75.Da}

\maketitle

The central limit theorem (CLT) for sums of independent random
variables 
\cite{gnedenko_1} plays a
fundamental role in statistical physics. This theorem is essential for
the construction of equilibrium statistical mechanics \cite{khinchin_1},
underlies the description of Brownian diffusion
\cite{gardiner_1}, and provides
justification to numerical approaches like Monte Carlo methods. 
According to it, the
probability density functions (PDF) of the sums are Gaussian when the single variable
PDF's have finite second moment.  
If on the other hand these PDF's have long-range
tails determining the divergence of the second moment, the
L\'evy-Gnedenko limit theorem states that the sums are L\'evy
distributed \cite{gnedenko_1,levy_1}.
There are however many situations, 
like critical phenomena in statistical 
systems \cite{kadanoff_1},  
financial time series \cite{mandelbrot_1}, anomalous transport \cite{bouchaud_1}, 
and protein dynamics \cite{kou_1},
where the presence of strong correlations
leads to non-Gaussian PDF's obeying
anomalous scaling with finite second moment \cite{brown_1}.
Understanding how correlations determine anomalous scaling and 
universality is still a challenge in general, at least outside 
equilibrium statistical mechanics. Indeed, in this context
renormalization group (RG) methods 
opened the way to
a probabilistic interpretation of scaling and universality in critical
phenomena \cite{lasinio_1}. 
RG transformations for effective Hamiltonians
provide a framework for the discussion
of critical scaling in cases when, due to the lack of
independence, the simple factorization of
individual variable characteristic functions (CF) on which the CLT
is based, does not hold. This framework
requires new and more complicated forms of limit theorems and stability
criteria \cite{lasinio_1}.
In view of the key role played by the CLT
in many fields, it is legitimate to ask if some form of 
CF's factorization helps in discussing the correlated case. This could
allow us to establish parallels between
the treatments of independent and strongly dependent variables. 

In this Rapid Communication we show that
a nonstandard factorization of summand variable CF's allows one to
construct the CF of the sum consistent with the assumption of
asymptotic anomalous scaling. 
This factorization is at the basis of a novel CLT 
and, under further
conditions, allows one to reconstruct the correlations of the asymptotic
process. 

In many physical situations, as one considers the sum
$X\equiv\sum_{i=1}^N X_i$ of stochastic variables $X_i$ with values $x_i$
in the real axis,
it is observed that for large $N$  the
PDF of $X$, $p_X(x)$, asymptotically obeys a simple scaling:   
\begin{equation}
N^{D} p_{X}(x)\sim g\left(\frac{x}{N^D}\right),
\label{scaling_eq}
\end{equation}
where $g$ is a scaling function and $D$ is a
scaling exponent. The scaling is anomalous if 
$g$ is not a Gaussian function or $D \neq 1/2$. As appropriate 
in most physical applications, we  
consider cases in which $p_{X}$ has finite
second moment.
The $X_i$'s and $X$ could be, respectively, the spins
and the total magnetization of a critical ferromagnetic system.
They could also represent the hour by hour increments
and the total variation of a return in a financial time series
which is sampled on intervals of $N$ hours.
Self-similarity is implied by Eq. (\ref{scaling_eq}) since 
plots of $N^D p_{X}$ vs $x/N^D$ at different $N$ asymptotically 
collapse onto the same curve $g$. 
To make this idea more precise, one can consider the normalized sum
$Y\equiv\sum_{i=1}^N X_i/N^D$, and its PDF
$p_{Y}(y)\equiv p_N(y)$. 
From Eq. (\ref{scaling_eq}) follows then, in the limit $N\to\infty$,
\begin{equation}
p_N(y)\sim g(y).
\label{scaling_eq_2}
\end{equation}
For the CF of $p_N$, 
$\tilde p_N(k)\equiv\int_{-\infty}^{+\infty} \exp(i k y)p_N(y) d x$,
Eq. (\ref{scaling_eq_2}) reads
\begin{equation}
\tilde p_N(k)\sim\tilde g(k),
\label{scaling_eq_fou}
\end{equation}
where $\tilde g$ is the Fourier transform (FT) of $g$
($\tilde p_N(0)=\tilde g(0)=1$).
We assume here that $p_{N}$ is even in $y$, so that 
$\tilde p_N(-k)=\tilde p_N^\ast(k)=\tilde p_N(k)$. 
Furthermore, $\tilde p_N(k)=1-\sigma^2k^2/2+O(k^4)$, where
the coefficient of $k^2$ is twice the second moment of $p_N$.  
Below, we choose units such that $\sigma^2/2=1$. 

If we consider  independent and 
identically distributed $X_i$'s, the CLT accounts for 
the asymptotic scaling in Eqs. (\ref{scaling_eq_2},\ref{scaling_eq_fou}) 
stating that it is not anomalous, i.e. it has $D=1/2$ {\it and}
$g$ Gaussian. 
Let us call $p_1$ the PDF 
of any individual $X_i$ and $\tilde p_1$ the corresponding CF.
By $N$-times convolution, one gets
\begin{equation}
\tilde p_N(k)=[\tilde p_1(k/N^{1/2})]^N.
\label{convolution}
\end{equation}
One can prove \cite{gnedenko_1} that 
$\tilde p_N$ becomes Gaussian (\mbox{$\sim\exp(-k^2)$}) at large $N$ 
for any $p_1$ with finite variance $\int_{-\infty}^{+\infty}
p_1(x) x^2 d x = 2$. Via inverse
FT this implies a Gaussian form for the asymptotic $p_N$ and $D=1/2$.
A key concept here
is that the limit PDF of the sum is stable, i.e. the sum
of two independent Gaussian distributed variables is
still Gaussian distributed.
This stability can be represented, e.g., by an invariance 
of $\tilde g$ under multiplication:
\begin{equation}
\tilde g(k/2^{1/2})\tilde g(k/2^{1/2})=\tilde g(k).
\label{invariance_eq}
\end{equation}
This functional relation directly follows
in the large $N$ limit from Eqs. (\ref{scaling_eq_fou},\ref{convolution}) 
and has \mbox{$\tilde g(k)= \exp(-k^{2})$} as
only possible solution. 

Here we
investigate the possibility of a generalization of the
multiplicative structure in Eq. (\ref{invariance_eq})
through the following steps:
(i) We assume the existence of a $\tilde g$ and a $D$
characterizing a given form of asymptotic anomalous scaling; 
(ii) We then introduce, in terms of $\tilde g$ and $D$ themselves, a
generalized multiplication $\otimes$ such that the identity
\begin{equation}
\tilde g(k/2^D)\otimes\tilde g(k/2^D)=\tilde g(k);
\label{invariance_eq_gen}
\end{equation}
holds; (iii) Eventually, we apply this generalized multiplication
$\otimes$ to CF's  different from $\tilde g$ in order to prove a CLT
implying the existence of a wide class of correlated processes
asymptotically behaving consistently with the scaling specified by
$\tilde g$ and $D$. 

We consider
scaling functions with the additional property  
of $\tilde g$ being strictly monotonic in $[0,+\infty)$ \cite{ising_2d}, 
which in our context also implies 
$0<\tilde g(k)\leq 1\;\forall k\in\mathbb R$. 
For $a_1,a_2\in(0,1]$, the generalized multiplication
allowing satisfaction of Eq. (\ref{invariance_eq_gen}) is 
\begin{equation}
a_1\otimes a_2\equiv
\tilde g
\left(
\left\{
\left[\tilde g^{-1}(a_1)\right]^{\frac{1}{D}}+
\left[\tilde g^{-1}(a_2)\right]^{\frac{1}{D}}
\right\}^{D}
\right),
\label{gen_multiplication}
\end{equation}
where $\tilde g^{-1}$ is the inverse of $\tilde g$ in $[0,+\infty)$.
One can easily verify that $a_1\otimes a_2\in(0,1]$, $a_1\otimes1=a_1$, and
that $\otimes$ is associative and
commutative \cite{kaniadakis_1}.
Eq. (\ref{invariance_eq_gen}) is recovered by putting 
$a_1=a_2=\tilde g(k/2^D)$. 
It is important to remark that 
if $\tilde g$ is Gaussian and $D=1/2$
the $\otimes$
multiplication reduces to the ordinary one.
The consideration of $a_1\neq a_2$ in
Eq. (\ref{gen_multiplication}) is clearly not needed to recover Eq. (6),
but becomes of crucial importance to determine 
joint probabilities for partial sums of the $X_i$'s
compatible with the anomalous scaling of the total sum
\cite{preparation}. 

One can further establish
a precise correspondence between this generalized multiplication 
and the ordinary one.
Once fixed $\tilde g$ and $D$, 
let us consider the mapping \mbox{${\cal M}_{\tilde g,D}:(0,1]\to(0,1]$} defined as
\mbox{${\cal M}_{\tilde g,D}(\cdot)\equiv 
\exp\left(-\left[\tilde g^{-1}(\cdot)\right]^{1/D}\right)$}
and its inverse
\mbox{${\cal M}_{\tilde g,D}^{-1}(\cdot)\equiv 
\tilde g\left(\left[-\ln(\cdot)\right]^D\right)$}.
Eq. (\ref{invariance_eq_gen}) can then be rewritten as:
\begin{eqnarray}
&{\cal M}_{\tilde g,D}^{-1}\left\{
{\cal M}_{\tilde g,D}\left[\tilde g(k/2^D)\right]\cdot
{\cal M}_{\tilde g,D}\left[\tilde g(k/2^D)\right]
\right\}=&\nonumber\\
&=
{\cal M}_{\tilde g,D}^{-1}\left\{
{\cal M}_{\tilde g,D}\left[\tilde g(k)\right]
\right\},&
\label{iso_eq}
\end{eqnarray}
which exemplifies the fact that 
${\cal M}_{\tilde g,D}$
establishes an isomorphism between the generalized and the ordinary
multiplications.
A key consequence
is that 
$\widehat{g}\equiv{\cal M}_{\tilde g,D}(\tilde g)$
obeys a condition of the form (\ref{invariance_eq}) 
with the exponent
$1/2$ replaced by $D$. This is the well known 
L\'evy-Gnedenko stability
condition for independent random variables, which has 
the singular L\'evy CF, 
$\exp\left(-|k|^{1/D}\right)$, as solution \cite{gnedenko_1,levy_1}.
Consistently, of course,   
$\widehat g(k)=\exp\left(-|k|^{1/D}\right)$.
Notice that
the L\'evy stable $\widehat{g}$ looses the meaning 
of CF for $D<1/2$,
because the corresponding PDF ceases to be positive definite.
Here this limitation does not apply, since the inverse FT of 
$\widehat{g}$ does not represent a PDF.

According to the L\'evy-Gnedenko limit theorem \cite{gnedenko_1}, 
the L\'evy stable
CF is approached in the $N \to \infty$ limit
for the sum of $N$ independent variables
whose individual CF has the same 
leading singularity $\sim |k|^{1/D}$ at $k=0$. This circumstance 
and the above mapping
suggest to look at the counterpart
of such convergence process in the space of correlated
PDF's.
In analogy with the independent case 
(Eq. (\ref{convolution})), 
we can indeed construct
the CF of the sum of $N$ 
correlated variables, 
starting from a single-variable CF $\tilde p_1$, 
but replacing the ordinary multiplication with the generalized one, 
as specified by the chosen $\tilde g$ and $D$. 
As before, $\tilde p_1$ is assumed to be regular and to generate 
a finite second moment, but in general will not coincide with $\tilde g$.
If we pose 
$\widehat{p}_1\equiv{\cal M}_{\tilde g,D}(\tilde p_1)$
this function is
singular at $k=0$: $\widehat{p}_1=1-|k|^{1/D}+O(|k|^{2/D})$.
Hence, by the L\'evy-Gnedenko limit theorem,
$[\widehat p_1(k/N^D)]^N\sim \widehat g(k)=\exp(-|k|^{1/D})$ 
for $N\to\infty$ \cite{gnedenko_1,levy_1}. 
The above isomorphism guarantees then that
\cite{homeomorphism} 
\begin{equation}
\left[\tilde p_1\left(\frac{k}{N^D}\right)\right]^{\otimes N}\equiv
\underbrace{\tilde p_1\left(\frac{k}{N^D}\right)\otimes\cdots\otimes
\tilde p_1\left(\frac{k}{N^D}\right)}_{N\;{\rm terms}}
\sim\tilde g(k)
\label{new_clt}
\end{equation}
for $N\to\infty$ and for any $p_1$ with finite variance $\sigma^2=2$.
Eq. (\ref{new_clt}) follows from the
fact that 
${\cal M}_{\tilde g,D}^{-1}(\widehat g)=\tilde g$ and expresses a CLT for
general $\tilde g$ and $D$. 
Starting from a single variable PDF $p_1$, the iterated
generalized multiplication of its CF 
yields the CF for the sum of the variables in a process where the
$X_i$'s are correlated. 
In force of the CLT, this process leads asymptotically
to the universal anomalous scaling specified by
$\tilde g $ and $D$.

\begin{figure}
\includegraphics[width=0.49\columnwidth]{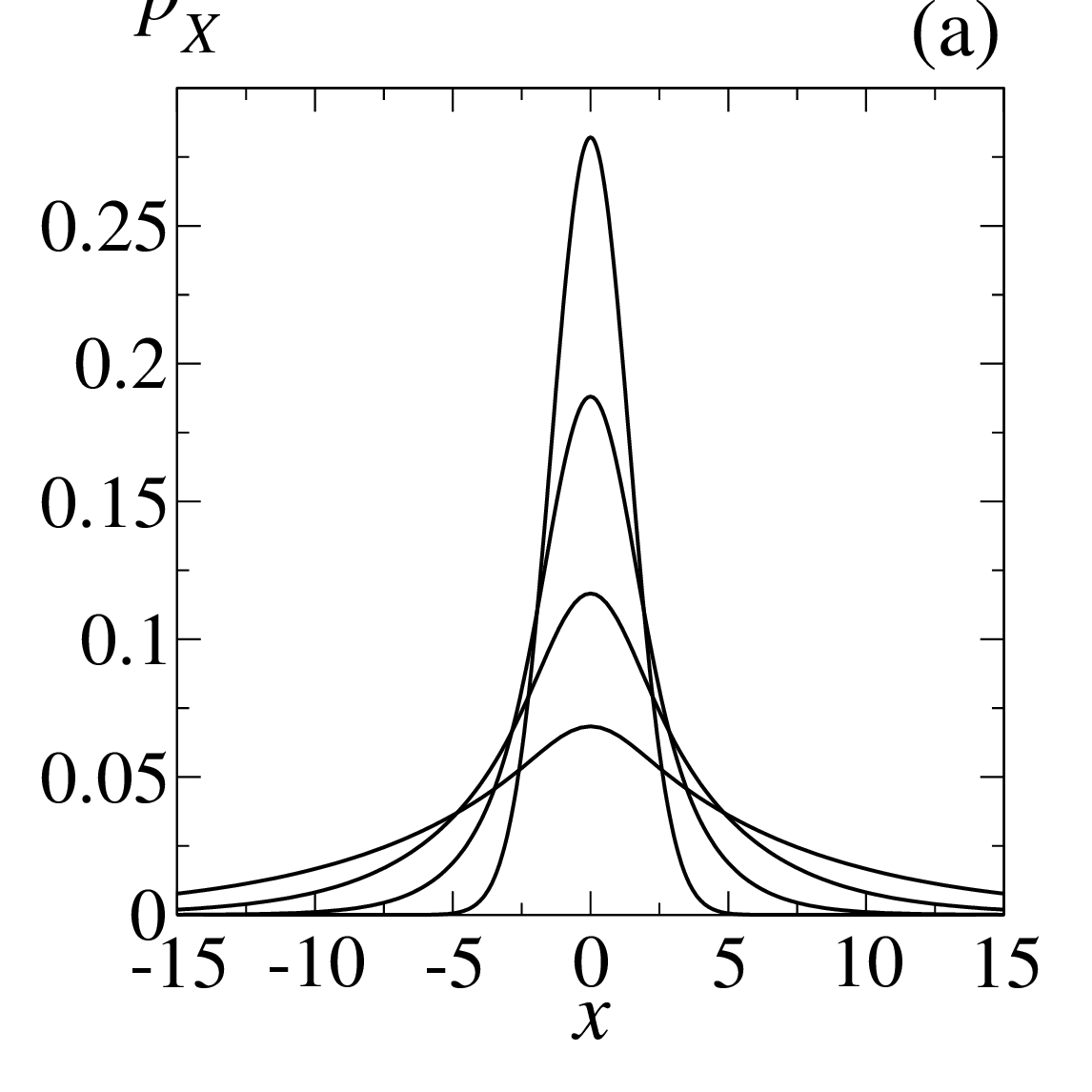}
\includegraphics[width=0.49\columnwidth]{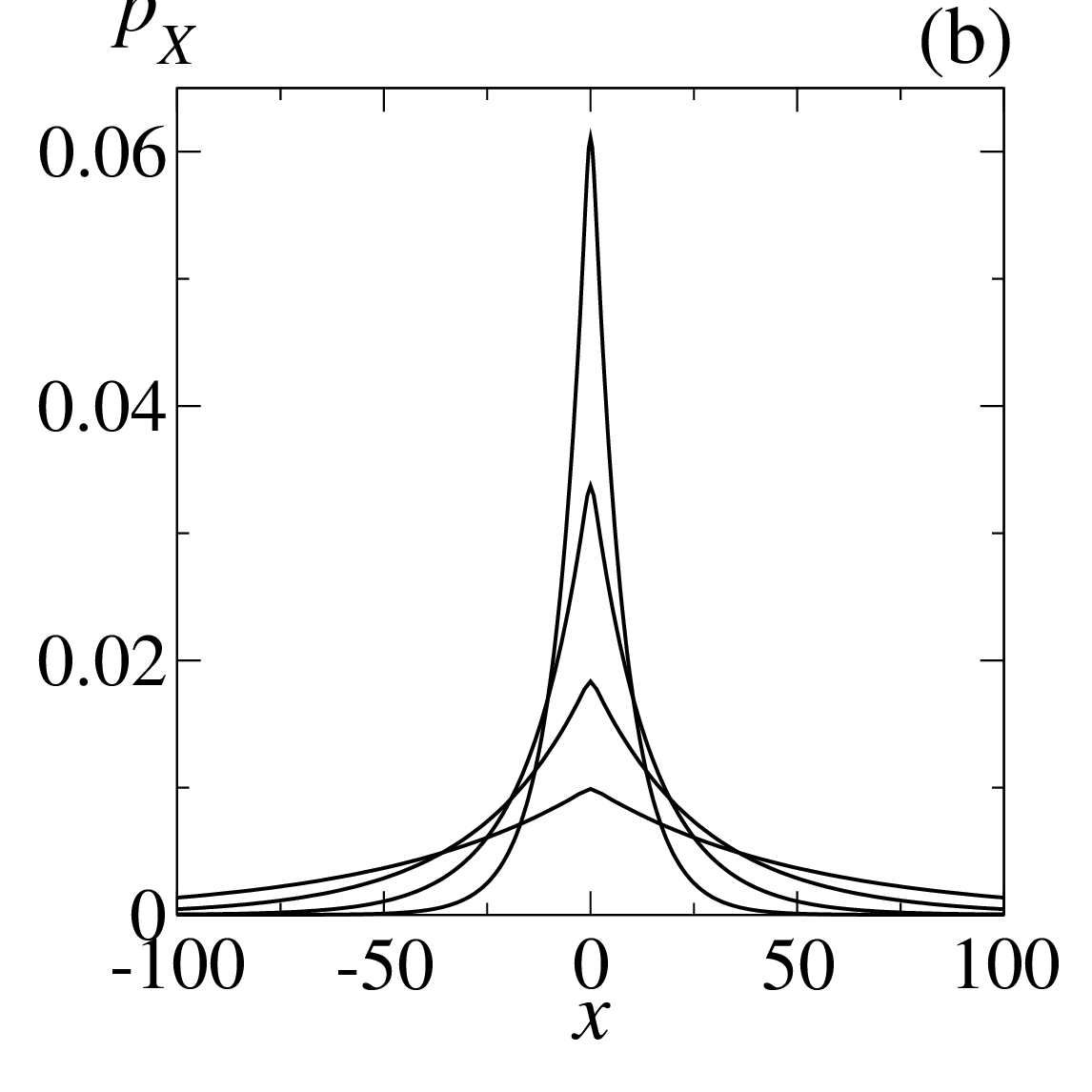}\\
\vspace{0.25cm}
\includegraphics[width=0.49\columnwidth]{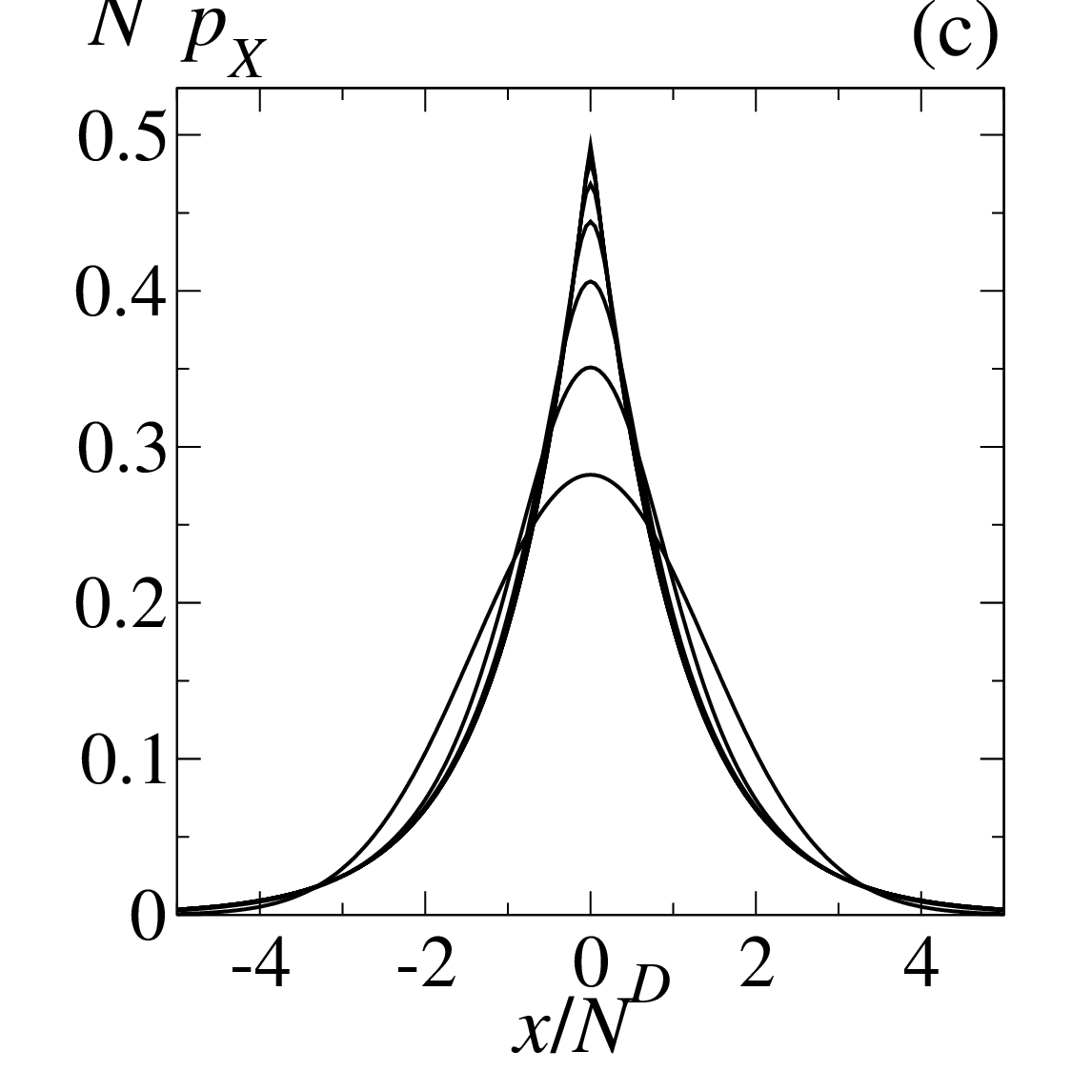}
\includegraphics[width=0.49\columnwidth]{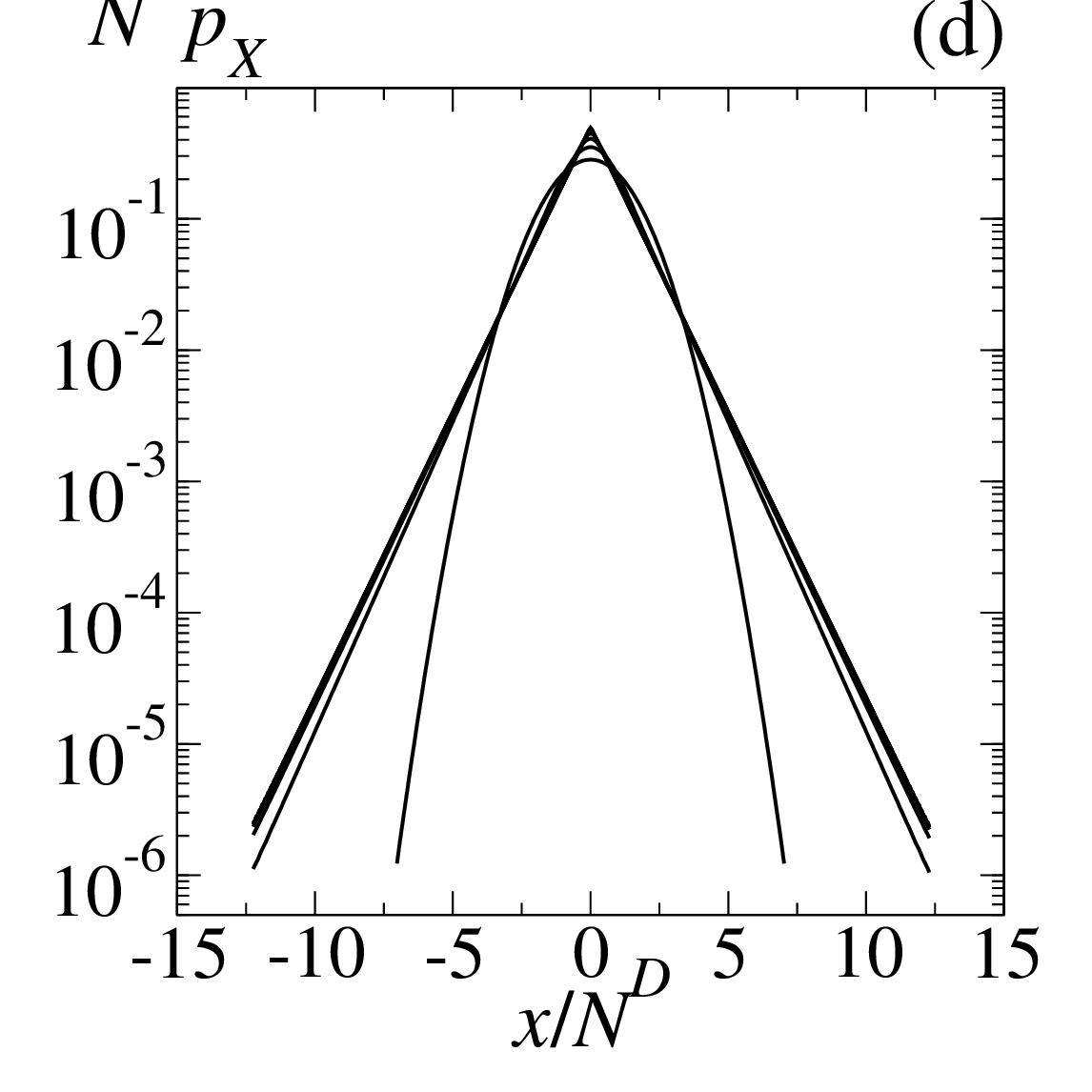}
\caption{
  Gaussian 
  $p_1$ and $D=0.9$.
  Plot of $p_X$ for $N=2^k$, $k=0,1,2,3$ (a), and
  $k=4,5,6,7$ (b). 
  As $N$ increases, the central peak of $p_X$ decreases.
  The rescaled plots for $k=0,1,\ldots,7$ have increasing
  peaks and reveal convergence for large $N$: (c) and (d). 
}
\label{fig_g_D0p9}
\end{figure}

\begin{figure}
\includegraphics[width=0.49\columnwidth]{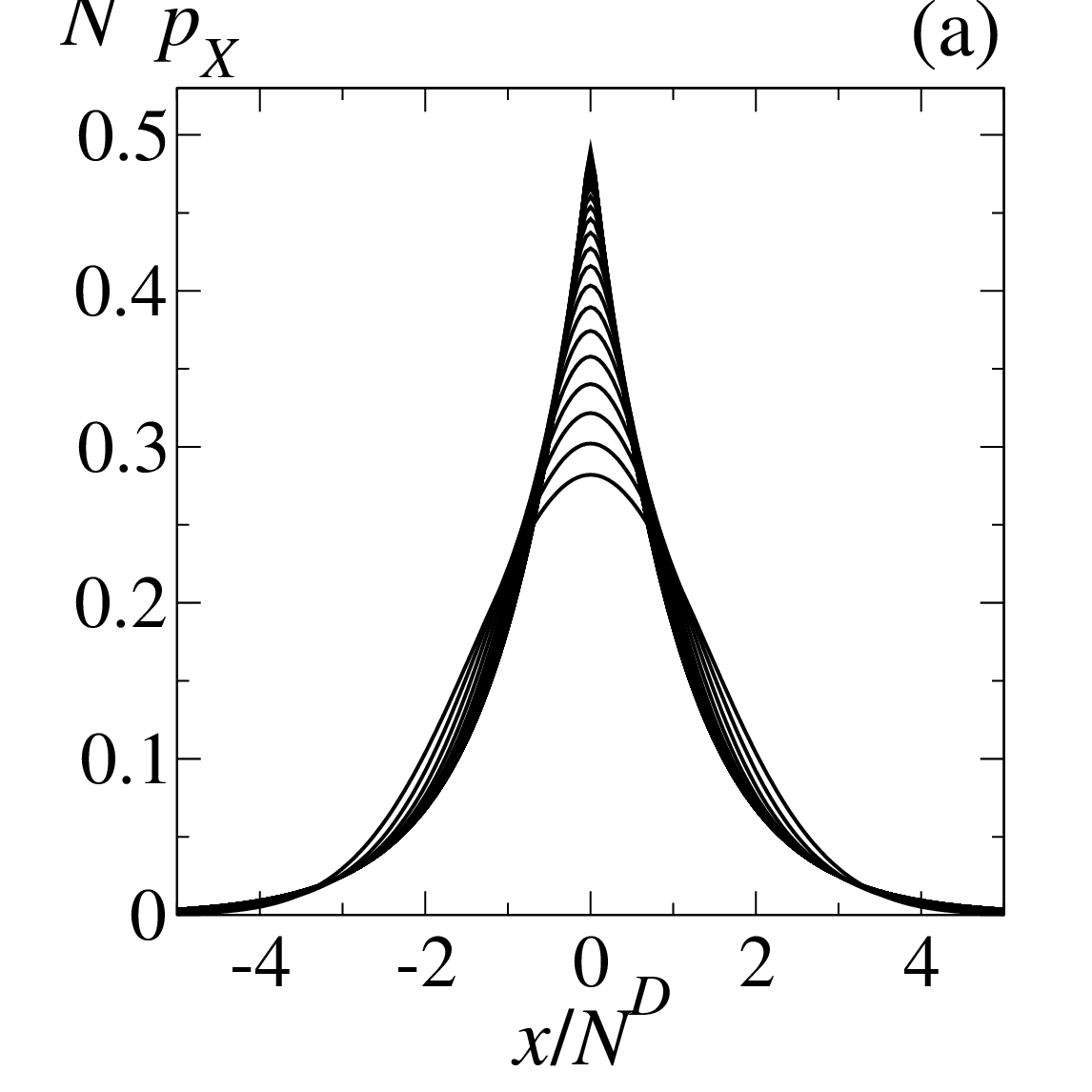}
\includegraphics[width=0.49\columnwidth]{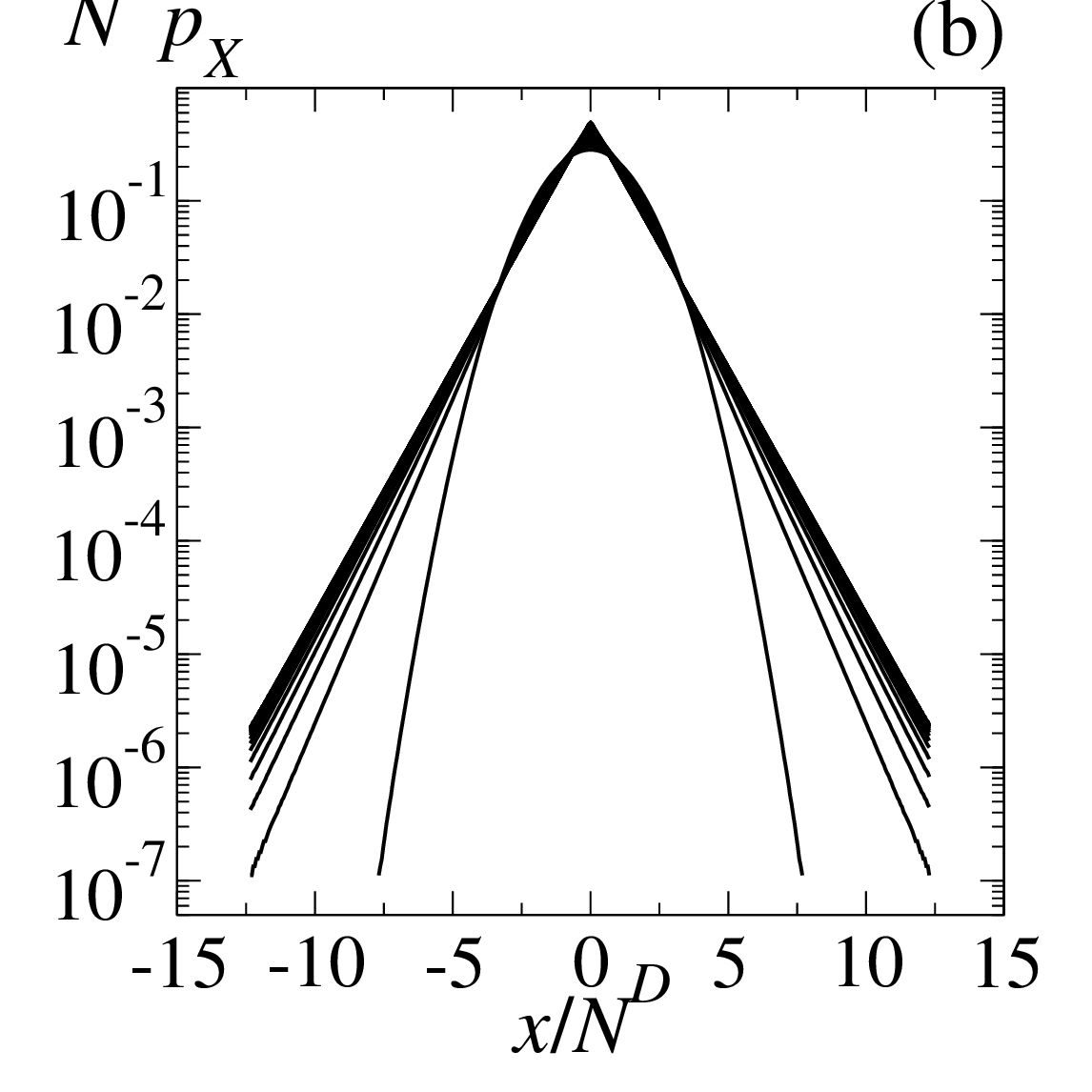}
\caption{
  Rescaled plots of $p_X$ with $D=0.25$, $N=2^k$, $k=0,1,\ldots,22$: (a) and (b). 
  Comparison with Fig. \ref{fig_g_D0p9} indicates slower convergence 
  when $D$ is smaller.
}  
\label{fig_g_D0p25}
\end{figure}

The validity of Eq. (\ref{new_clt}), does not require $D>1/2$ because
again the inverse FT of $\widehat{p}_1$ is not constrained to remain 
positive. However, other positivity requirements can 
pose limits on the choice of $p_1$. Indeed, there is no
guarantee that, if $\tilde p_1$ is a CF, 
$\tilde p_1 \otimes \tilde p_1$ will also be,
in general. 
Since positivity control is a hard mathematical issue
\cite{gnedenko_1,giraud_1}, we addressed it numerically 
by analyzing the convergence process in 
Eq. (\ref{new_clt}) for
several $\tilde g$'s and $\tilde p_1$'s.
We verified that 
as long as $\tilde p_1$ has the same general properties assumed
for $\tilde g$, 
$p_N(y)\equiv(1/2\pi)\int_{-\infty}^{+\infty} 
\exp(-i k y)[\tilde p_1(k/N^D)]^{\otimes N}d k$
remains positive definite for any $N$.
For illustration we report the results for the case 
$\tilde g(k)=1/(1+k^2)$, i.e., $g(y)=\exp(-|y|)/2$ and 
$\tilde g^{-1}(a)=-\sqrt{1/a-1}$ for $a\in(0,1]$.  
Fig. \ref{fig_g_D0p9} and \ref{fig_g_D0p25}
show the evolution of $p_{X}(x)$ under the generalized
multiplications of the single variable CF for a Gaussian 
$p_1(x)=\exp(-x^2/4)/\sqrt{4\pi}$ and, respectively, $D=0.9$ and
$D=0.25$. In general, larger $D$'s imply faster
convergence to the fixed-point. However, after a sufficient
number of iterations, all the collapses we checked are almost perfect.  
One may wonder if Eq. (\ref{new_clt}) remains valid for
more general forms of $p_1$. 
A first extension of the above results can be obtained by
considering  
single variable
PDF's with two symmetric peaks, which, e.g., could be relevant for
magnetic or diffusive phenomena. 
In this case 
$\tilde p_1$
is not strictly positive anymore, so that a continuation of the
generalized multiplication to negative values is required. 
One can indeed find a continuation that preserves the isomorphism
with the ordinary multiplication \cite{preparation}.
We verified \cite{preparation} that while Eq. (\ref{new_clt})
remains valid asymptotically, for this new class of $\tilde p_1$'s
positivity problems of the iterated PDF's can arise during the initial
stages of the convergence process. 

In all examples, 
only the constraint $\sigma^2=2$ and, possibly, positivity
requirements, pose limitations on the domain of attraction of the
stable PDF.
This universality, typical of the CLT, is a
consequence of the multiplicative structure in Eq. (\ref{new_clt}).  
Indeed, in Eq. (\ref{new_clt}) normalization ($\tilde p_N(0)=1^{\otimes N}=1$),
centering ($\int_{-\infty}^{+\infty}y p_N(y)d y=0$), and 
variance ($\int_{-\infty}^{+\infty}y^2 p_N(y)d y=\sigma^2=2$) are
conserved,
as in  the independent case. 
Thus, the trajectory described by $p_N$ in function space differs
substantially from a typical  
RG flow, in which the variance is not conserved and relevant
scaling fields determine the critical surface \cite{lasinio_1}. Here, 
relevant fields are not present \cite{eigenvalues}
and the result in Eq. (\ref{new_clt}) identifies at least a subset of
the universality domain of the assumed
asymptotic anomalous scaling specified by $g$ and $D$.
A further feature of our findings
is that 
the choice of $g$ does not imply a selection on
admissible values of $D$, and vice-versa. 
This appears consistent with the variety of different anomalous
scaling functions and exponents observed in natural phenomena 
\cite{tsallis}. 

A basic issue is that of identifying an explicit mechanism by
which correlations are introduced by the $\otimes$ multiplication.  
Let us consider the normalized partial sums
$Y_1=\sum_{i=1}^{N/2}X_i/(N/2)^D$ and
$Y_2=\sum_{i=N/2+1}^{N}X_i/(N/2)^D$. The correlations between
$Y_1$ and $Y_2$ are fully specified once their joint 
PDF $p_N^{(2)}(y_1,y_2)$ is given. The knowledge of
$p_N$ alone does not allow to determine $p_N^{(2)}$ in general
since many $p_N^{(2)}$ are such to satisfy the obvious
condition
\mbox{$p_N(y)=\int_{-\infty}^{+\infty}p_N^{(2)}(y_1,y_2)\;
\delta(y-y_1-y_2)\;d y_1d y_2$}. So, many different correlation
patterns are compatible with the anomalous scaling of $p_N$.
Eq. (\ref{new_clt}) asymptotically fixes
$\tilde p_N^{(2)}(k/2^D,k/2^D)\sim\tilde g(k/2^D)\otimes g(k/2^D)$,
where $\tilde p_N^{(2)}$ is the FT of $p_N^{(2)}$. In the independent case
this last result would hold with $\tilde g$ Gaussian, $D=1/2$ and the standard
multiplication replacing $\otimes$. Furthermore, $\tilde p_N^{(2)}(k_1/2^{1/2},k_2/2^{1/2})
\sim \tilde g(k_1/2^{1/2}) \tilde g(k_2/2^{1/2})$ would clearly hold
in that case also for $k_1 \neq k_2$, so that
$\tilde p_N^{(2)}$ would be fully specified in terms of $\tilde g$. 
It is natural to ask if,
under suitable assumptions, an analogous factorization
of $\tilde p_N^{(2)}$ with $k_1 \neq k_2$ holds in terms of
the $\otimes$ multiplication also 
in the correlated case. This property would
imply the possibility of expressing the correlations
determining the anomalous scaling in terms of $p_N$ alone.
It can be shown \cite{preparation} that such a factorization is indeed possible 
if additional symmetries of $p_N^{(2)}$ are assumed, like the
vanishing of linear correlations between $Y_1$ and
$Y_2$,
\mbox{$\int_{-\infty}^{+\infty} y_1 y_2 \;p_N^{(2)}(y_1,y_2)\;
d y_1d y_2=0$}. 
This vanishing does not hold for other stochastic processes
possessing anomalous scaling 
considered in the literature, like for example the fractional Brownian
motion \cite{mandelbrot_2}.
Because of market efficiency the vanishing of linear correlations
characterizes, e.g., financial time series, where $p_N$
is the PDF of the normalized return of an index in time $N$. 
For such series we were able to show \cite{preparation} 
that the asymptotic form
of $\tilde p_N^{(2)}$ 
can be uniquely determined
starting from $p_N$ and using a $\otimes$ multiplication.
The agreement of the theoretical predictions with the
empirically sampled $p_N^{(2)}$ is quite remarkable \cite{preparation}.
Thus, the generalized multiplication operation defined above
is also a key for the full characterization of a 
relevant class of stochastic evolution processes.

One can also establish a connection between the
present CLT and anomalous diffusion. 
Let us consider a single variable PDF of the form  
$p_1(x)=[\delta(x-\Delta)+\delta(x+\Delta)]/2$,
and define a time $t\equiv N\tau$. Here, releasing the 
condition $\sigma^2=2$, $\Delta$ and $\tau$ are, respectively, the
space and time span of random steps, and $t$ is the time
at which the $N$-th step occurs.
In the continuum limit $N\to\infty$, $\tau\to0$,
$\Delta\to0$, such that $t$ and 
${\cal D}^{1/2D}\equiv\Delta^{1/D}/\tau$ remain finite, one
recovers \cite{preparation}
$\lim_{N\to\infty}[\tilde p_1(k)]^{\otimes N}\equiv\tilde p(k,t)=
{\cal M}_{\tilde g,D}^{-1}[\widehat p(k,t)]$,
where $\widehat p(k,t)$ satisfies the standard L\'evy
diffusion equation \cite{hughes_1}
\begin{equation}
\frac{\partial\widehat p(k,t)}{\partial t}
=-\frac{{\cal D}^{1/2D}|k|^{1/D}}{2^{1/2D}}\widehat p(k,t).
\end{equation}
Assuming $p(x,0)=\delta(x)$, one gets the solution 
$\widehat p(k,t)=\exp(-{\cal D}^{1/2D}|k|^{1/D}t/2^{1/2D})$,
which corresponds to 
$\tilde p(k,t)=
\tilde g({\cal D}^{1/2} k t^D)=
1-{\cal D}k^2t^{2D}/2+O({\cal D}^2k^4t^{4D})$. 
Hence, $\langle x^2\rangle(t)={\cal D}t^{2D}$
\cite{preparation}. 
Thus, correlated sub- ($D<1/2$) and super- ($D>1/2$) diffusive
solutions can be obtained through our
mapping from the propagator of the uncorrelated L\'evy diffusion
Eq. (10). This enables the description of 
the evolution towards the asymptotic anomalous diffusion regime
(analogous to Figs. \ref{fig_g_D0p9},\ref{fig_g_D0p25}) 
without introducing a broad distribution of waiting times elapsing
between successive steps as it is done in the continuous
time random walk approach \cite{bouchaud_1,shlesinger_1}.  

In summary, assuming anomalous scaling 
(Eqs. (\ref{scaling_eq},\ref{scaling_eq_2}))
we have constructed a multiplicative functional identity 
for the CF of the asymptotic sum of strongly correlated random
variables which allowed the definition of a generalized
multiplication. 
An isomorphism between this multiplication and the ordinary
one leads to establish a CLT in which the anomalous scaling represents 
the asymptotic limit.
Thus, for a given asymptotic anomalous scaling form
we have characterized a large class of processes falling in the
corresponding universality domain. 
In particular cases our strategy also allows a full 
determination of the joint probabilities,
and thus of the correlations of partial sums of the random variables
\cite{preparation}.  
In the context of stochastic processes, the
presence of correlations implies that past events have an influence on
the future behavior. 
The knowledge of the joint probability of consecutive events (like
$Y_1$ and $Y_2$) hence
entails a predictive power which has been recently 
exploited in finance \cite{preparation}.

{\bf Acknowledgments:}
We thank M. Baiesi for collaboration and 
J.R. Banavar, H. Fogedby, G. Jona-Lasinio, F. Leyvraz, 
D. Mukamel, E. Orlandini, M. Paczuski, C. Tebaldi, H. Touchette for
discussions.


\begin{thebibliography}{99}
\bibitem{ising_2d}
With additional technical effort \cite{preparation} our results can be
extended to more general cases, like the magnetization PDF of the
critical 2D Ising model 
\cite{lasinio_1}.
\bibitem{homeomorphism}
The rigorous 
conditions under which our correspondence becomes a homeomorphism
are not discussed here.
\bibitem{eigenvalues}
For Gaussian $g$ and arbitrary $D$ one can linearize the transformation
in Eq. (\ref{new_clt}) at the asymptotic fixed point and compute its eigenvalues
explicitly \cite{preparation}.
\end{thebibliography}
\end{document}